# Effects of internal asymmetry on interface curvatures and outer drags determining the oriented shift of the eccentric globules

Jingtao Wang[1*], Genmiao Xu[1], Pan Hu[1], Jing Guan[2**]

[1] School of Chemical Engineering and Technology, Tianjin University, Tianjin, 300072, P. R.China

[2] School of Science, Tianjin University, Tianjin, 300072, P. R.China

*wjingtao928@tju.edu.cn, **guanjing@tju.edu.cn

The physical mechanism of the oriented shift and inverse of eccentric globules in a modest extensional flow are investigated in this paper. Through this work, a shift of the globule, which is driven mainly by the asymmetric interfacial curvature, not by the outer drag, is disclosed. The asymmetric layout of the daughter droplet leads to the asymmetric drags from the continuous phase and the asymmetric deformation of the globule with different interface curvatures. As the inner droplet has both enhancing and suppressing effect on the globule deformation, the interface curvatures will vary when changing the relative size and location of the inner droplet. This curvature difference results in the asymmetric pressure distribution and circulation inside the globule. Eventually, the interaction of the inner driving force (pressure differences) and the outer drags causes the oriented shift and inverse of the globule. The shift direction is affected not only by the structural asymmetry parameter ε (eccentricity), but also by some flow features such as the capillary number. The results obtained here might enlighten potential applications for the movement of soft globules driven by curvature differences.

## 1. Introduction

As complex soft particles, multiple-emulsion globules could have skillfully-designed internal structures (Wang , Zhang , & Chu, 2014). Since they might have multiple engulfed independent droplets and sequentially broad potentials, they have drawn much attention recently in many different fields such as the drug delivery, food and cosmetics industry et. al. (Wang , Wang , & Han, 2011). Most of time, these complex globules are stored and delivered in liquid phases, and they always deform under the action of the flow shear. Thus, it is natural and important to study their rheological behaviors carefully.

Up to now, many works have been done to investigate the rheology of multiple-emulsion globules, especially for concentric double emulsions (CDE), in different flow systems by using experiments (Chen, Li, Shum, Stone, & Weitz, 2011; Li , Chen , & Stone , 2011; Chen, Liu, Zhang, & Zhao, 2015; Chen, Gao, Zhang, & Zhao, 2016) and numerical simulations (Stone, & Leal ,1990; Kan et al.,1998; Toose et al.,1999; Smith et al.,2004; Chen, Liu, Zhang, & Zhao, 2015). The controlled release (Chen, Li, Shum, Stone, & Weitz, 2011; Li , Chen , & Stone , 2011) and split (Chen, Gao, Zhang, & Zhao, 2016) of double-emulsion globules under the flow shear had been done experimentally in three-dimensional microchannels assembled by capillary tubes. Besides these experiments, the rheological behaviors of multiple-emulsion globules have been studied through numerical methods by many researchers. As early as in 1990, Stone and Leal (1990) have presented two different breakup mechanisms of CDE globule by studying their deformation and breakup in infinite extensional flows through a boundary element method. Chen et al. (Chen, Liu, Zhang, & Zhao, 2015) investigated the effects of the inner droplet of CDE on its deformation in shear flows. They demonstrated that there are two coexisting enhancing and suppressing effects of the inner droplet on the deformation of the globule when the deformation process reaches an equilibrium state. Moreover, the rheological behaviors of the multiple-emulsion globule in straight and constriction tubes have been investigated by using numerical simulations. (Zhou, Yue, & Feng, 2008; Tao, Song, Liu, & Wang, 2013; Wang, Li, Wang, & Guan, 2014). Through a spectral boundary element method, Wang et al. (2014) simulated the translation process of double-emulsion globules containing two unequal daughter droplets in a constriction tube. It is very interesting that the required maximum pressure drop is relatively lower when the initial location of the bigger daughter droplet is in the front of the globule, which means that the globule is easier to pass the constriction in this way.

Lately, the rheological behaviors of multiple-emulsion globules with complex asymmetric internal structures and the oriented movement of their inner droplets have drawn much attention of some researchers due to their potential applications in the controlled release of the globule insertion. Wang and coworkers (Wang, Liu, Han, & Guan, 2013a; Wang, Liu, Han, &Guan, 2013b.; Wang, Wang, Tai, &Guan, 2016) designed asymmetric multiple-emulsion globules with three layers. In the $2^{nd}$ layer, there is one big daughter droplet (DD) which has asymmetric internal structures. Due to the structural asymmetry in the $3^{rd}$ layer, the daughter droplet will shift in a special direction, which might cause the oriented contact of the outmost interface of the globule and that of DD. Thus, the globule might break up and release its insertion orientably. Moreover, Wang et al. (2016) disclosed that the shift due to the internal asymmetry is controlled not only by the asymmetric structures but also by flow features such as the viscosity ratio and capillary number. By changing these factors, they could inverse the shift direction. As for the eccentric compound globule, Qu and Wang (2012) have investigated its rheological behaviors in a planar extensional flow. They reported that the eccentric compound droplet will shift away from its original location in the same direction as the moving direction of the inner droplet. The speed of the globule is proportional to the distance between two centers of globule and the inner droplet, i.e., proportional to the eccentricity. However, all these works have not investigated and analyzed the phenomena and mechanical mechanism of oriented shift of the globule deeply, and could not present the physical explanations of the shift reverse.

In addition, the asymmetric rheological process of a simple droplet in asymmetric extensional flows has been studied by Wang and coworkers (Wang, &Yu, 2015; Yu, Zheng, Jin, & Wang, 2016). In their works, the droplet is symmetric, but the outer flow surrounding the drop is asymmetric. When the capillary number is small, the droplet could not break up and always shift in the same direction in which the droplet suffers bigger outer drags from the continuous phase. However, for a multiple-emulsion globule with an internal asymmetry in an extensional flow under modest capillary numbers, as the asymmetry is from the interior of the globule, its shift might be more complex and might not be caused by only the outer drag.

## 2. Brief introduction of the numerical method

In this paper, the rheological behaviors of a double-emulsion globule (Fig. 1b and 1d) with one eccentric daughter droplet (DD) in an axisymmetric cross-like microfluidic device (Fig. 1a and 1c) is studied numerically. Based on this investigation, the effect of internal asymmetry on the interfacial curvature, the inner pressure distribution and the outer drags of the globule are disclosed. Here, we employ a spectral boundary element method (Dimitrakopoulos, & Wang, 2007; Wang, Liu, Han, & Guan, 2013a) based on Fig. 1a and 1b. The validation of this numerical method has been well done in our previous papers (Wang, Liu, Han, & Guan, 2013b). According to Wang et. al.'s work (2016), the capillary number and the parameter to define the asymmetry are the two most important factors affecting the oriented shift. Thus, two critical parameters are studied in this paper. The first is still the capillary number Ca, and the second is a dimensionless number, eccentricity, which is defined by $\varepsilon=d_R/(r_{MR}-r_R)$. Here, $r_{MR}$ and $r_R$ are the radii of the globule and the daughter droplet, respectively, and $d_R$ is the eccentric distances of DD. By increasing $d_R$ or $r_R$, the eccentricity of the complex globule could be enhanced.

All physical quantities are dimensionless here and they are reduced by some scale. For example, the length scale is the radius $R_0=1$ of the outlet channel. Time is scaled by the flow time scale $G_0^{-1}$, where $G_0$ is the shear rate at the wall of the outlet when the volume flow rate $Q$ is $2R_0^2/3$. The pressure scale is $\mu_0 G_0$, the viscosity and the interface tension are scaled by $\mu_0$ (the viscosity $\mu$ of the continuous phase) and $\mu_0 R_0 G_0$, respectively. Initially, all droplets are spherical due to surface tensions. The radius of the globule $r_{MR}=0.5R_0$ and that of DD $r_R=0.1R_0$. Thus, the ratio $\Phi$ of the inner droplet size to the globule is fixed to 4%. The daughter droplet is always located on x-axis and $d_R$ will be changed to adjust the eccentricity $\varepsilon$. The viscosity ratio $\lambda$ of the dispersed phase (DP) to the continuous phase (CP) is fixed to 1.2. As for Ca=$\mu U/\gamma$, as the viscosity of CP is fixed $\mu=1$ and the average velocity of CP at outlets is fixed $U=1$, the interfacial tension $\gamma$ will be changed to adjust the value of Ca.

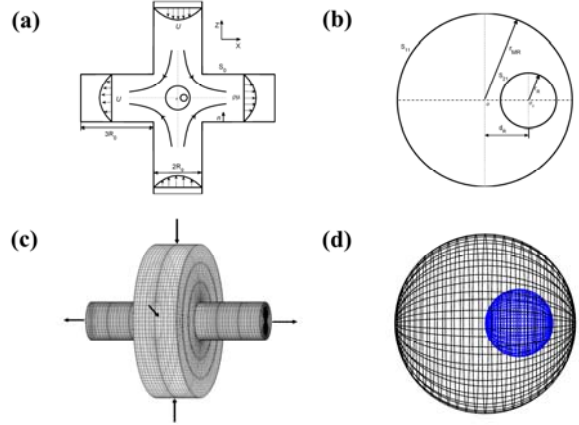

Fig. 1. (a) and (b) The illustration of the 2-dimensional cross-like microfluidic device and eccentric double-emulsion globule. (c) and (d) The illustration of the equivalent axisymmetric cross-like microfluidic device and axisymmetric multiple-emulsion globule.

## 3. Results

In the investigated flow system, when taking the globule interface as the boundary, the continuous phase surrounding the globule is the outer flow; the dispersed phase between the interfaces of the globule and the daughter droplet is the inner flow. Thus, the entire flow field is divided into two regions: the outer flow which could generate drags to the globule, and the inner circulation which could move the mass center of the globule. Apparently, the asymmetric layout of the daughter droplet would cause an aymmetric pressure distribution and result in an oriented circulation inside the globule (See Fig. 2). Thus, the globule might shift in some specific direction (to the left or right), and the direction is determined by both the internal asymmetry and some parameters changing flow features such as Ca. As shown in Fig. 2, under the same Ca=0.2, double-emulsion globules with different initial eccentricities have opposite pressure distributions and inner circullations, and further have the opposite shift directions.

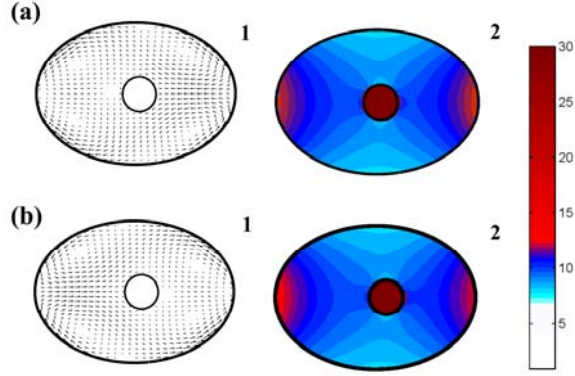

Fig. 2. (a) When the initial ε=0.375 and Ca=0.2, the eccentric globule will shift to the left due to the asymmetric internal circulation and pressure distribution. (b) When the initial ε=0.750 and Ca=0.2, the same globule will shift to the right due to the asymmetric internal circulation and pressure distribution.

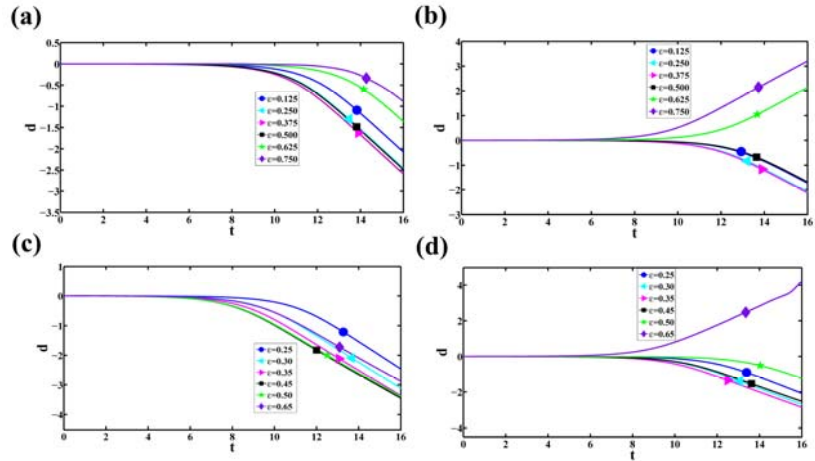

Fig. 3 The displacement of the mass center of the globules with one eccentric daughter droplet along with time at various initial eccentricities and capillary numbers (Ca =0.16 for figure a and c; Ca =0.20 for figure b and d). For figure 3a and 3b, $r_R$=0.1$R_0$ is fixed and the initial ε changes through the variation of the eccentric distances $d_R$. For figure 3c and 3d, $d_R$=0.1$R_0$ is fixed and the initial ε changes through the variation of the inner droplet radius $r_R$.

The displacement $d$ of the mass center of the eccentric globules along with time at various capillary numbers and initial eccentricities are investigated and shown in Fig. 3. When the initial ε is changed from 0.125 to 0.750 through the variation of the eccentric distances $d_R$, the capillary number is 0.16 for Fig. 3a and is 0.20 for Fig. 3b. Comparing Fig. 3a to Fig. 3b, it is obvious that the shift behaviors of the same eccentric globule under two capillary numbers are different. As for Fig. 3a, the eccentric globule intends to shift to the left when the initial eccentricities variy from 0.125 to 0.750. However, although the globule always shifts to the left in this case, the shift speed is not monotonically increasing. From ε=0.125 to 0.375, the speed of shift to the left increases; however, from ε=0.375 to 0.750, the speed decreases continuously. As for Fig. 3b, the eccentric globule might shift either to the left or to the right when the initial ε is in the range from 0.125 to 0.750. From ε=0.125 to 0.375, the globule shifts to the left and its speed increases; from ε=0.375 to 0.500 the globule shifts to the left but its speed decreases. When the initial eccentricities are 0.625 and 0.750, the shift directions of the globule change to the right and the speeds increase. When the initial ε is changed from 0.25 to 0.65 through the variation of the inner droplet radius $r_R$, the capillary number is 0.16 for Fig. 3c and is 0.20 for Fig. 3d. Comparing Fig. 3c to Fig. 3d, it also could see that the shift behaviors are different for the same eccentric globule under various capillary numbers. As for Fig. 3c, the eccentric globule always tends to move to the left when the initial eccentricity varies in the range of 0.25 to 0.65. However, its shift speed is also not monotonically increasing: from ε=0.25 to 0.45, the shift speed increases; from ε=0.45 to 0.65, the speed decreases. As for Fig. 3d, the eccentric globule could shift either to the left or to the right. When the initial ε varies from 0.25 to 0.35, the globule shifts to the left and its speed increases; from ε=0.35 to 0.50 the globule shifts to the left but its speed decreases continuously. When the initial ε is increased to 0.65, the globule inverses its shift direction and moves to the right. From the above analysis, it could assert that the relatively small initial eccentricity and its appropriate increase will benefit the shift to the left; however, when beyond a limit, the increment of the eccentricity is beneficial to the left shift no longer, and might benefit the shift to the right when the initial ε is big enough. In addition, from Fig. 3, we also could know that the relatively large Ca is beneficial to the right shift.

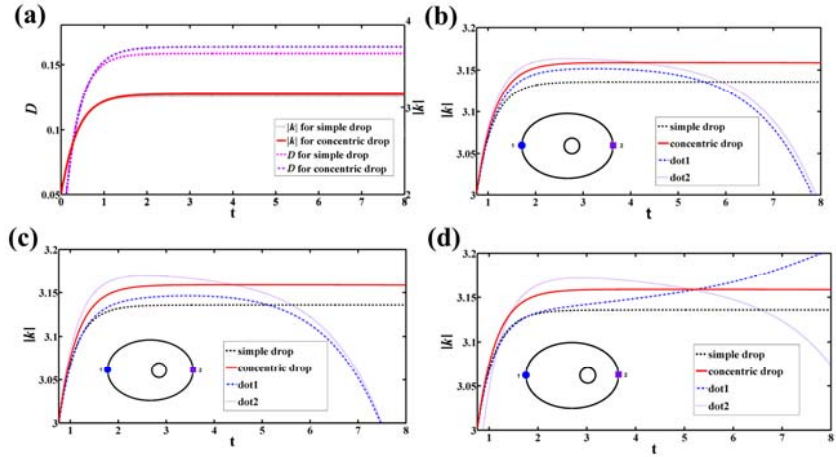

Fig. 4 The deformation parameter *D* and interfacial curvatures (absolute values |*k*|) at two ends of the globules *versus* time *t* for the simple drop, concentric globule and eccentric globule under Ca=0.2. (a) *D* and |*k*| *versus t* for the simple drop and concentric globule. |*k*| *versus t* for eccentric globule with initial ε=0.125 (b), ε=0.375 (c) and ε=0.750 (d).

In order to explain the phenomena shown in Fig. 3, Fig. 3b has been analyzed further and the obtained results are shown in Fig. 4, Fig. 5 and Fig. 6, respectively. At first, the interfacial curvatures *k* at the right and left end of the deformed globule are investigated and shown in Fig. 4.

When the inner droplet is exactly staying at the center of the globule, we have a concentric globule whose eccentricity ε is zero. As shown in Fig. 4a, when the deformation process reaches an equilibrium, both the deformation parameter $D$ defined according to the reference papers (Dimitrakopoulos, & Wang, 2007; Wang, Han, & Yu, 2012) and the curvature for the concentric globule are bigger than those for the simple droplet, which means the inner droplet will enhance the deformation at equilibrium. As for the eccentric globule, actually there is no deformation at equilibrium due to the asymmetric internal structure. Nevertheless, as it is well known that the inner circulation has four symmetric eddies when the deformation of a concentric globule reaches the equilibrium, the deformation process for the eccentric globule could still be divided into two stages according to the pattern of the inner circulation: before the formation of the circulation with four eddies (BF) and after that (AF). In the AF stage, according to displacement of the globule mass center, the deformation could also be separated into two stages: unclear (AFUD) and clear displacement (AFCD). Thus, the curvature curves for eccentric globules in Fig. 4b-4d have three stages: BF stage (t=0 to about t=1), AFUD stage (about t=1 to about t=4), and AFCD stage (about t>4). As our interest is to investigate the physical cause of the oriented shift, we will concentrate on BF and AFUD stage since AFCD stage is the result not the origin. As shown in Fig. 4b-4d, due to the globule eccentricity, the curvature at the left end (dot 1) is different from that at the right end (dot 2). In order to show the effects of the inner droplet of the eccentric globules on the interfacial curvatures, the curvature curves for the simple droplet and the concentric globule are also shown in Fig. 4b-4d as references. In AFUD stage, it is obvious that the curvature at dot 2 is larger than that for the concentric globule since the inner droplet is closer to dot 2 and thus has a stronger enhancing effect. Also, the curvature at dot 1 is less than that for the concentric globule, but still larger than that for the simple droplet. This is reasonable since the inner droplet in eccentric globules is farther to dot 1 than that in the concentric globule and thus has a weaker enhancing effect. However, in BF stage, the situation is just opposite. Although the difference might be very small when the eccentricity is low, it could still assert that the inner droplet has suppressing effects for the globule deformation in BF stage. It could see that in this stage the curvature at dot 1 is larger than that at dot 2 at the same moment, which is much clearer when the eccentricity is high.

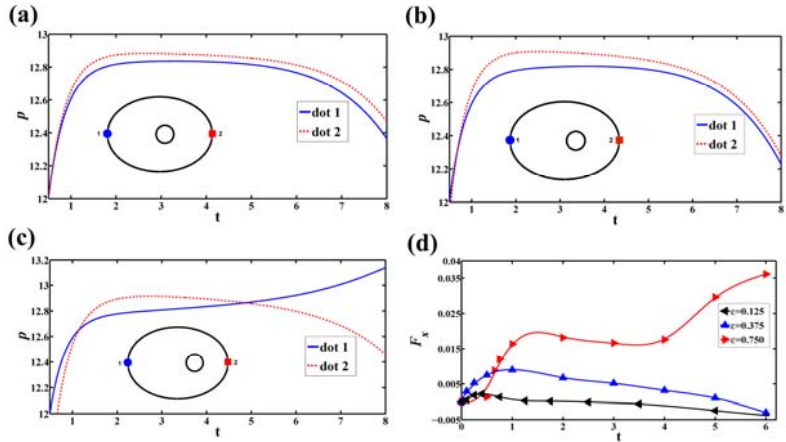

Fig. 5 The inner pressure $p$ at two ends of the eccentric globules *versus* time $t$ for the initial eccentricity ε=0.125 (a), ε=0.375 (b) and ε=0.750 (c) under Ca=0.2. (d) The x component of the total drag force $F_x$ *versus* time $t$ for eccentric globules with different initial eccentricities.

The enhancing and suppressing effects of the inner droplet on the interfacial curvatures could also be seen in the pressure curves (Fig. 5a-5c) of two ends of the eccentric globules. For the three different initial eccentricities ε=0.125 (Fig. 5a), ε=0.375 (Fig. 5b) and ε=0.750 (Fig. 5c), the pressures at dot 1 are always higher than those at dot 2 in BF stage, which could cause the right shift of the globule; and in AFUD stage, the pressures at dot 1 are always lower than those at dot 2, which could cause the left shift. However, in AFCD stage, the pressures at dot 1 are always lower for ε=0.125 and ε=0.375, which is consistent to the obvious shift of the globule to the left; as for ε=0.750, the situation is just opposite at the most time, which is consistent to the obvious shift of the globule to the right. In order to disclose the mechanical mechanism behind the oriented shift, the curves of the x component $F_x$ of the total drag forces for three initial eccentricities are shown in Fig. 5d. Since the flow system is totally symmetric in y direction, the y component $F_y$ is always zero. When ε=0.125 and ε=0.375, initially, values of $F_x$ are positive (which means that the force points to the right) and increase along with time. Then, when $F_x$ reaches a maximum value, the curves will decline along with time, and in AFCD stage they might have negative values. When ε=0.750, values of $F_x$ are always positive and much higher than those for ε=0.125 and ε=0.375 at most of time. In BF stage, the curve of $F_x$ ascends rapidly; in AFUD stage, values of $F_x$ are relatively quite large but do not change very much; in AFCD stage, the curve of $F_x$ ascends along with time again. From the above analysis, it could see that the outer drag is generally stronger to the half of the eccentric globule in which the inner droplet is staying. Furthermore, since the inner droplets always stay in the right half of the globule for the cases studied in this paper, it could be asserted that the outer drags always tend to pull the globule to the right in both BF and AFUD stage no matter how big the initial eccentricities are.

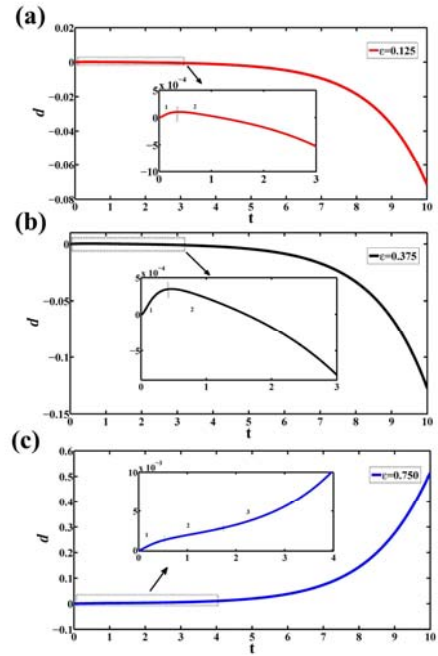

Fig. 6 The displacement *d* of the mass center of the eccentric globules *versus* time *t* for various initial eccentricities under Ca=0.2. (a) ε=0.125, (b) ε=0.375 and (c) ε=0.750.

In order to see the effects of inner pressure differences and outer drag forces on the shift of the eccentric globules further, the time-evolution curves of the mass center of the globules are shown in Fig. 6, and the curves for both BF and AFUD stage are particularly magnified. When $\varepsilon=0.125$ and $\varepsilon=0.375$ (see Fig. 6a and 6b), initially, the globules shift to the right a little bit, which is the actively-combined action of the inner pressure differences and outer drag forces. Then, the curves of $d$ descend continuously, which means that the globules begin to shift to the left. At this moment, it could see from Fig. 5a, 5b and 5d that the outer drags pull the globule to the right but the inner pressure differences move the globule to the left. Thus, it could assert that the left shifts of the eccentric globules are caused by the inner pressure differences, i.e., by the interfacial curvature differences. When $\varepsilon=0.750$ (see Fig. 6c), the $d$ curve is always positive and ascends continuously although the ascending rates in different stages change a little bit. As shown in the insertion of Fig. 6c, the ascending rate in stage one is relatively high since both the inner pressure difference and the outer drag are positive to the right shift in this stage; in stage two, the outer drag is quite big and positive but the inner pressure difference (which is diminishing quickly, see Fig. 5c) is negative to right shift, thus, the slope of $d$ curve in this stage decreases a little. In stage three, the inner pressure difference and the outer drag are both positive to the right shift again. Thus, we could assert that the right shift of the eccentric globule is caused by the outer drag.

Although the investigation in Fig. 4-6 only focuses on the cases shown in Fig. 3b (in which the initial $\varepsilon$ changes through the variation of the eccentric distances $d_R$), the corresponding results for the cases shown in Fig. 3d (in which the initial $\varepsilon$ changes through the variation of the inner droplet radius $r_R$) are similar and could make the same conclusion. In addition, that the larger capillary number is favorable to the right shift is also reasonable since the larger Ca means the higher shear rate and the stronger outer drag.

## 4. Conclusions

In summary, the mechanical mechanisms of the oriented shift and inverse of eccentric complex globules in a modest extensional flow have been investigated in this paper. According to our common sense, generally the movement of a globule is driven by the asymmetric outer drags. However, a globule shift driven by the asymmetric interfacial curvature is disclosed in this work. As the inner droplet has both enhancing and suppressing effect on the deformation of the globule, the asymmetric layout of the inner droplet leads to the asymmetric deformation of the globule with different interface curvatures at two ends. This curvature difference results in the asymmetric pressure distribution and circulation inside the globule, which could drive the globule shifting in some direction by changing its mass center. In addition, the internal asymmetry also results in the asymmetric drags from the continuous phase, which could drive the globule shifting too. The higher the asymmetry is, the larger the outer drag is. The interaction between the outer drag and the inner driving force (inner pressure difference) causes the oriented shift and inverse of the globule eventually. Thus, the shift direction is affected not only by the structural asymmetry parameter $\varepsilon$ (eccentricity), but also by some flow features such as the capillary number. Changing these factors might cause the variation of the shift direction. When the initial $\varepsilon$ and Ca are relatively small, the globule is mainly driven by the asymmetric interfacial curvatures and shift to the left; when $\varepsilon$ and Ca are relatively big, the globule is mainly driven by the outer drags and shift to the right. The results obtained in this paper might have significant potential applications for the curvature-driving movement of soft globules.

**Acknowledgements**

Supported by National Natural Science Foundation of China (21376162 and 21576185).